\documentclass[a4paper,11pt]{article}
\pdfoutput=1 

\usepackage{jcappub} 

\usepackage[T1]{fontenc} 

\usepackage{amsfonts}
\usepackage{amsmath}
\usepackage{amssymb}
\usepackage{graphicx,color}
\usepackage{float}
\usepackage{hyperref}
\usepackage{subfigure}
\usepackage{dcolumn}
\usepackage{soul}
\usepackage{verbatim}
\usepackage{bm}  
\usepackage{booktabs}
		
\title{Thermal enhancement of inflationary magnetic fields} 

\author[a]{Arjun Berera,}
\author[a,b,1]{Suddhasattwa Brahma,\note{Corresponding author.}}
\author[a]{Zizang Qiu,}
\author[c]{Rudnei O. Ramos}

\affiliation[a]{Higgs Centre for Theoretical Physics, School of Physics and Astronomy, University of Edinburgh, Edinburgh EH9 3FD, UK}
\affiliation[b]{Physics and Applied Mathematics Unit, Indian Statistical Institute, 203 B.T. Road, Kolkata 700108, India}

\affiliation[c]{Departamento de F\'{\i}sica Te\'orica, Universidade do Estado do
  Rio de Janeiro, 20550-013 Rio de Janeiro, RJ, Brazil }

\emailAdd{ab@ed.ac.uk}
\emailAdd{suddhasattwa.brahma@gmail.com}
\emailAdd{Zizang.Qiu@ed.ac.uk}
\emailAdd{rudnei@uerj.br}

\abstract{
We investigate primordial magnetogenesis by assuming the gauge field is 
prepared in a thermal state during inflation rather than the standard 
Bunch-Davies vacuum. The temperature $\mathcal{T}$ introduces a physical 
scale that breaks conformal invariance at the level of the state while preserving 
the standard Maxwell action. This modification results in a {\it dissipative boost} that alters the magnetic energy density scaling from $a^{-4}$ to $a^{-3}$, resulting in a present-day magnetic field $B_0$ enhancement that can potentially range from about $10^{8}$ to $10^{12}$ on cosmological scales. While this toy model alone does not satisfy observational lower bounds, it demonstrates that thermal initial conditions can significantly mitigate the conformal obstruction. Our results suggest that embedding this mechanism within a fully dynamical warm inflation framework, where dissipation continuously maintains the thermal bath, provides a highly promising path towards successfully realizing a minimal model of inflationary magnetogenesis without the need to invoke non-minimal couplings, anomalous background dynamics or nonlinear extensions of electrodynamics.
}

\begin{document}
\maketitle


\section{Introduction}

Magnetic fields are ubiquitous across the observable universe, and yet their 
physical origin remains a fundamental challenge in modern cosmology~\cite{Subramanian:2015lua}. Observations have confirmed their presence in 
planets, stars, galaxies \cite{1509.04522}, and clusters~\cite{2002ARA&A..40..319C}, with evidence even suggesting the existence of 
fields in the intergalactic medium (IGM)~\cite{2010Sci...328...73N} that have 
coherence lengths extending to megaparsec (Mpc) scales. While the galactic 
dynamo mechanism is often invoked to amplify pre-existing fields~\cite{Widrow:2002ud}, it inevitably requires a primordial `seed' field to initiate the process. Astrophysical mechanisms, such as the Biermann battery during 
structure formation \cite{Kulsrud:1996km} or outflows from Active Galactic 
Nuclei (AGN)~\cite{Furlanetto:2001gx}, typically lack the causal reach to 
generate the large-scale coherence observed in voids. Conversely, inflation 
naturally provides a mechanism for generating super-horizon correlations from a 
single causal region~\cite{Guth:1980zm}, potentially explaining the presence of 
magnetic fields in high-redshift systems and low-density regions~\cite{Turner:1987bw, Durrer:2013pga,Widrow:2011hs,Barrow:2011ic}.

Despite this theoretical appeal, concrete realizations of inflationary 
magnetogenesis face severe hurdles. In standard Maxwell electrodynamics, the 
action is conformally invariant in a flat Friedmann-Lemaître-Robertson-Walker 
(FLRW) background; consequently, vacuum fluctuations of the gauge field $A_\mu$ 
are diluted by the expansion of the universe as $1/a^4$. To generate viable 
seed fields, models typically break conformal invariance by non-minimally 
coupling the gauge field to the inflaton, \textit{e.g.} $I(\phi) F^{\mu\nu}F_{\mu\nu}$, or to the curvature, for example,  $f(R) F^{\mu\nu}F_{\mu\nu}$, where $R$ is the Ricci scalar while $\phi$ denotes the inflaton field\footnote{Both these couplings are equivalent during vanilla slow-roll inflation \cite{Teuscher:2025jhq}.}. However, these approaches are constrained by the `strong-coupling' and `backreaction' problems~\cite{Subramanian:2015lua, 2016JCAP...03..010G}. The former occurs when the effective gauge coupling becomes non-perturbative during inflation, while the latter refers to the electromagnetic energy density becoming large enough to disrupt the inflationary background. Quantitative analysis suggests a ``no-go'' type bound for two-derivative models, restricting the energy scale of inflation to values barely compatible with Big Bang Nucleosynthesis (BBN) constraints~\cite{2016JCAP...03..010G}.

In this work, we explore an alternative path: preserving the minimal structure 
of the electromagnetic action while modifying the initial state of the gauge 
field from the standard Bunch-Davies vacuum to a thermal state. This setup is 
naturally motivated by the paradigm of warm inflation~\cite{Berera:1995ie}, 
where dissipative dynamics lead to the continuous production of a radiation 
bath. Unlike cold inflation, where radiation is diluted as $a^{-4}$, the 
continuous energy transfer from the inflaton in the warm scenario can maintain 
a near-constant physical temperature $\mathcal{T}$ throughout the expansion.
This thermal bath introduces a preferred physical scale $\mathcal{T}$, which 
breaks the conformal symmetry of the system at the state level rather than at 
the level of the action. This leads to a modification of the magnetic power 
spectrum, where the effective redshifting of the field energy density is 
reduced from $a^{-4}$ to $a^{-3}$. We treat this as a toy model of a thermal 
state in a de Sitter background within the test-field limit, distinguishing it 
from thermal fluctuations in a plasma~\cite{Tajima1992ApJ390309, Lemoine:1995fh}. While a fully realistic model requires a quasi-de Sitter treatment and consistent dynamical coupling, this work focuses on identifying and isolating the primary physical mechanism for such enhancement due to a thermal state. 

This paper elucidates the primary physics that generates the magnetic
fields, and future work will embed this toy model into a full warm inflationary scenario with all its associated extra details. In anticipation of embedding this mechanism in a fully dynamical warm inflation model, where the thermal state maintains its temperature due to continuous levels of high dissipation, we term the enhancement in the magnetic field amplitude as \textit{`dissipative boost'}. Moreover, it is also conceivable that the gauge field particles are in a thermal
state during inflation, created by a warm inflation type
particle production dynamics, yet at the same time the inflaton
field is not interacting very strongly with this or other thermalized
fields, and hence it remains in its ground state producing cold inflation
type vacuum fluctuation. Thus, there are also possibilities
for our scenario to be realized in such a hybrid of warm and cold inflationary
dynamics. Finally, the thermal state will also enhance the magnitude of the primordial electric field, as we will show below, which can further induce magnetic fields in a reheating phase when conductivity is low \cite{Kobayashi:2019uqs}.  All such details with regard to specific scenarios we leave for future work.

Note that the conformal symmetry is kept intact at the level of the action, and the thermal state is not invariant under the symmetry group due to the introduction of a preferred scale parametrized by $\mathcal{T}$. Since this is not a global symmetry, Goldstone's theorem does not apply. Instead, the symmetry breaking by the thermal state---which is a mixed state---manifests as temperature-dependent modifications in the spectrum of the gauge field. 

The remainder of this paper is organized as follows: In Section~\ref{sec2}, we review the existing constraints on inflationary magnetogenesis. In Section~\ref{sec3}, we present our main results concerning the thermal enhancement of the spectrum. Finally, Section~\ref{sec4} provides our concluding remarks and prospects for future work.

\section{Bounds on Inflationary Magnetogenesis}
\label{sec2}

Let us start with the standard generic approach to model-building for inflationary magnetogenesis to recapture the current constraints. The Maxwell action for the electromagnetic gauge field $A_\mu$ is given by
\begin{eqnarray}
	S=-\int {\rm d}^4 x\,\sqrt{-g}\, \left[\frac{1}{16\pi}g^{\mu\alpha}g^{\nu\beta}F_{\mu\nu}F_{\alpha\beta} + \ldots\right]\,,
\end{eqnarray}
where $F_{\mu\nu}$ is the standard field strength and $\ldots$ represents the coupling of $A_\mu$ to other terms. We have assumed a conformally flat FLRW metric of the form ${\rm d}s^2 = a^2(\tau) [-{\rm d}\tau^2+ {\rm d}{\bf x}^2]$, where the scale factor takes the form $a(\tau) = -1/(H_{\rm inf}\tau)$ for standard slow-roll inflation in the leading order. Here $\tau$ is the conformal time as given in the metric and $H_{\rm inf}$ is the inflaton scale. 

Choosing the Coulomb gauge, $A_0({\bf x},\tau) = {\bf \nabla}. {\bf A}({\bf x},\tau) = 0$, Maxwell's equation for the transverse part of the vector potential reads
\begin{eqnarray}\label{Maxwell's-Eqn}
	A_i'' - \nabla^2 A_i = 0\,
\end{eqnarray}
where here $i$ runs over the spatial indices $(1,2,3)$, $\nabla^2\equiv\partial_{\bf x}^2$ is the spatial Laplacian over the comoving coordinates and primes denote derivatives with respect to the conformal time $\tau$. This implies standard plane-wave solutions for the mode function for the gauge field once it is expanded as follows:
\begin{equation}
	A^i(\bold{x},\tau)= \sqrt{4\pi} \sum_{\lambda=1}^2 \int\frac{{\rm d}^3\bold{k}}{(2\pi)^3} \left[ \bold{e}_\lambda^i(\bold{k}) a_\lambda(\bold{k}) A(k,\tau)e^{i\bold{k}\cdot\bold{x}} + \bold{e}_\lambda^{i*}(\bold{k}) a^\dag_\lambda(\bold{k})A^*(k,\tau)e^{-i\bold{k}\cdot\bold{x}}\right]\,.
	\label{gauge-field-expansion}
\end{equation}
The index $\lambda$ runs from $1$ to $2$ representing the two physical transverse polarizations, and $\bold{e}_\lambda^i(\bold{k})$ are the spatially projected polarization vectors that satisfy the usual completeness relation $\sum_\lambda {e}_{i,\lambda}(\bold{k}) {e}_{j,\lambda}^{*}(\bold{k}) = \delta_{ij} - k_ik_j/{\bf k}^2$, where $k$ is the comoving wavenumber. Canonical quantization imposes the usual commutator between the creation and annihilation operators, namely,
\begin{equation}
	[a_\lambda(\bold{k}), a^\dag_{\lambda^\prime}(\bold{k}^\prime)]=(2\pi)^3\delta^3(\bold{k}-\bold{k}^\prime)\delta_{\lambda\lambda^\prime}\,,
\end{equation}
with all other commutators vanishing. The vacuum state is identified as the one annihilated by $a_\lambda(\bold{k})$, \textit{i.e.,} $a_\lambda(\bold{k})|0\rangle=0$. 

Substituting the expansion \eqref{gauge-field-expansion} into Maxwell's equation \eqref{Maxwell's-Eqn}, we find
\begin{eqnarray}
	A^{\prime\prime}_{k} + k^2A_{k}
 = 0\,.
\end{eqnarray}
This has the general plane-wave solution
\begin{eqnarray}
	A_{k}(\tau) = c_1(k) e^{-ik\tau} + c_2(k) e^{ik\tau}\,.
\end{eqnarray}
{}For inflationary magnetogenesis, one typically assumes the Bunch-Davies mode functions corresponding to the vacuum state defined above, which requires that
\begin{eqnarray}
	A_{{k}}(\tau) = \frac{1}{\sqrt{2k}} e^{-ik\tau}\,, \ \ \ \ {\rm as} \ \ \tau \rightarrow -\infty\,,
\end{eqnarray}
ensuring that all modes were sub-horizon in the infinite past and they were simply in the Minkowski vacuum. Given such initial conditions, one can fix $c_1=1/\sqrt{2k}$ and $c_2=0$. 

{}For a comoving observer with a $4$-velocity $u^\mu = (\frac{1}{a},0,0,0)$, where the temporal coordinate is the conformal time $\tau$, one can define the electric and magnetic fields as \cite{Subramanian:2015lua,Durrer:2013pga}
\begin{eqnarray}
	E_\mu = u^\nu F_{\mu\nu} \,, \ \ \ \ \ \ B_\mu = \frac{1}{2} \epsilon_{\mu\nu\rho} F^{\nu\rho}\,,
\end{eqnarray}
where $\epsilon_{\mu\nu\rho} = \varepsilon_{\mu\nu\rho\sigma} u^\sigma$, and $\varepsilon_{\mu\nu\rho\sigma}$ is the totally antisymmetric tensor corresponding to the usual Levi-Civita symbol. The convention we follow here is $\varepsilon_{0123}=-\sqrt{-g}$. With this definition, the spectral magnetic energy density can be written as 
\begin{eqnarray}\label{magnetic_power_spectrum}
	\frac{{\rm d} \rho_B}{{\rm d} \ln k} = \left(\frac{k^4}{2\pi^2 a^4}\right) \ k \ |A_{k}(\tau)|^2\,.
\end{eqnarray}
Hence, the magnetic power spectrum for the Bunch-Davies vacuum (with plane wave solutions) can be evaluated to be $k^4/(4\pi^2 a^4)$. 

A few years ago, Green and Kobayashi managed to find a model-independent way to put bounds on a broad class of inflationary magnetogenesis models in which the gauge field $A_\mu$ has the standard two-derivative kinetic term (to leading order) with a time-dependent prefactor $I(\tau)$ \cite{2016JCAP...03..010G}. We recap their main argument here for the convenience of the reader. The function $I$ was kept arbitrary subject to gauge invariance and reduction to standard Maxwell theory at the end of inflation. 
\begin{equation}
	S \supset \int {\rm d}\tau\,{\rm d}^3x\, \frac{I^2(\tau)}{2}\,A_i' A_i' + \cdots\,,
\end{equation}
where $I(\tau)$ is a dimensionless function with $I(\tau_f)\simeq 1$ so that the effective gauge coupling is given by $e_{\rm eff} = e / I$, with $\tau_f$ denoting the end of inflation. To ensure that the theory remains weakly-coupled throughout inflation, they required that
\begin{equation}
	I^2(\tau) \gtrsim 1\,.
\end{equation}
They imposed a backreaction constraint that the kinetic energy in the electromagnetic field never exceeds the background energy density to not disrupt inflation, namely,
\begin{equation}\label{Backreaction}
	\rho_{\rm kin}(\tau)\;\sim\;\int \frac{{\rm d}k}{k}\,\frac{k^3}{2\pi^2}\,\frac{I^2}{a^4}\,| A_k'|^2 \;\ll\; 3 M_{\rm Pl}^2 H_{\rm inf}^2\,.
\end{equation}
The change in the gauge field was quantified as
\begin{equation}
	| A_k(\tau_f)| \;\leq\; | A_k(\tau_i)| + \int_{\tau_i}^{\tau_f} {\rm d}\tau\,| A_k'(\tau)|\,,
\end{equation}
Note that the inequality comes from the complex nature of the gauge field as a function of a real variable $\tau$ \cite{2016JCAP...03..010G}. 
Together with the bound
\begin{equation}
	\frac{k^3}{2\pi^2}\,\frac{I^2}{a^4}\,| A_k'|^2 \;\ll\; 3 M_{\rm Pl}^2 H_{\rm inf}^2\,,
\end{equation}
one can derive that
\begin{equation}
	|A_k(\tau_f)| \;\lesssim\; \sqrt{6\pi}\,M_{\rm Pl}\,\frac{a_f}{k^{3/2}}\,.
\end{equation}
Given the magnetic power spectra is defined as  $P_B(\tau,k) \;\simeq\; \frac{k^5}{2\pi^2 a^4}\,|A_k|^2$ \eqref{magnetic_power_spectrum}, this implies an upper bound on the present-day magnetic spectrum:
\begin{equation}
	P_B(\tau_0,k)\;\lesssim\;(10^{-15}\,{\rm G})^2\left(\frac{k/a_0}{\mathrm{Mpc}^{-1}}\right)^2
	\left(\frac{H_{\rm reh}}{H_{\rm inf}}\right)^{1/3}
	\left(\frac{10^{-14}\,{\rm GeV}}{H_{\rm inf}}\right)\,,
\end{equation}
such that achieving $B_0\sim 10^{-15}\,{\rm G}$ on Mpc scales requires $H_{\rm inf}\lesssim 10^{-14}\,{\rm GeV}$. If a further constraint is that the gauge-field sector does not overproduce curvature perturbations, \textit{i.e.,}
\begin{equation}
	\zeta_A \;\sim\; \frac{1}{\epsilon}\,\frac{\rho_A}{\rho_{\rm inf}} \;\lesssim\; 10^{-5}
\end{equation}
is imposed, then this tightens the above bound to \cite{2016JCAP...03..010G}
\begin{equation}
	H_{\rm inf}\;\lesssim\;10^{-19}\,{\rm GeV}\,,
\end{equation}
which corresponds to reheating temperatures $T_{\rm reh}\lesssim 10^2\,{\rm MeV}$, only barely consistent with BBN bounds.

In order to produce large magnetic fields with a thermal state, we would need to identify the loophole in the above no-go result. A crucial assumption in the above argument made by the authors of \cite{2016JCAP...03..010G} is that $A_\mu$ is prepared in its vacuum state, or have classical growth sourced by some stochastic field variables. As we shall show in the next section, this assumption is explicitly violated by a thermal state and hence, in principle, we can circumvent their result.

\section{Thermal state for the electromagnetic field}
\label{sec3}

Physics in the real world typically come within a statistical state of particles, most commonly a thermalized state with corresponding finite temperature effects. In particular, non-equilibrium dynamics in the early universe may have placed the electromagnetic field in a thermal or near-thermal state. A concrete realization of this would be warm inflation which would produce radiation continuously, sourced by the inflaton energy density, in a near thermal state due to large dissipative effects. In the following, we will assume that there is some such mechanism which will put the electromagnetic field in a thermal state with a near constant physical temperature.

\subsection{Thermally Corrected Magnetic Energy Spectrum}

When the gauge field $A_\mu$ is in thermal equilibrium in a state with comoving temperature $T$, the mean occupation number, for a given comoving momentum mode $k$, is given by the Bose-Einstein distribution,
\begin{eqnarray}
	\langle n_k \rangle_\Omega = \frac{1}{e^{k/T} - 1}\,,
\end{eqnarray}
where we have denoted the thermal state as $|\Omega\rangle$. Since we will hold the physical temperature $\mathcal{T}$ constant during inflation, the comoving temperature $T(\tau) = a(\tau) \mathcal{T}$ increases exponentially during this period. Certainly, this assumption deserves some careful justification. 

This approximation of a near-constant physical temperature $\mathcal{T}$ during the inflationary epoch can be motivated most effectively in the context of the warm inflation paradigm~\cite{Berera:1995ie} 
(for reviews see~\cite{Berera:2008ar,Kamali:2023lzq}). In warm inflation, radiation production occurs concurrently with inflationary expansion.  Thus, warm inflation differentiates itself from the standard "cold" inflationary scenario, where the temperature drops exponentially as $\mathcal{T} \propto a^{-1}$. In this framework, 
the inflaton field $\phi$ is coupled to a bath of radiation through a dissipation term $\Upsilon$. The evolution of the radiation energy density $\rho_{\rm rad}$ is governed by the modified conservation equation,
\begin{equation}
    \dot{\rho}_{\rm rad} + 4H\rho_{\rm rad} = \Upsilon \dot{\phi}^2 \,.
\end{equation}
During the slow-roll phase, a quasi-stationary equilibrium is reached, where the production of radiation from the decaying inflaton compensates for the dilution caused by the expansion of space. In this steady-state regime, the radiation density remains nearly constant and, since $\rho_{\rm rad} \propto \mathcal{T}^4$ for a thermalized bath, the physical temperature $\mathcal{T}$ is effectively locked to a value determined by the dissipation strength and the Hubble rate,
\begin{equation}
    \mathcal{T} \approx \left( \frac{\Upsilon \dot{\phi}^2}{4H \sigma_{SB}} \right)^{1/4} \approx \text{constant}\,,
\end{equation}
where $\sigma_{SB}$ is the Stefan-Boltzmann constant. Consequently, while the comoving temperature $T = a\mathcal{T}$ grows exponentially, the local physical environment experienced by the electromagnetic field remains thermalized at a stable energy scale. This provides a self-consistent background for the thermal state $|\Omega\rangle$, allowing for a sustained enhancement of the magnetic power spectrum across the entire inflationary $e$-foldings. We verify this explicitly in the following. As mentioned before, since the root cause of sustaining such a thermal state is due to dissipation, we call this a \textit{dissipaive boost}.

The thermal state can be formally described as a squeezed state generated via 
a Bogoliubov transformation of the zero-temperature Bunch-Davies vacuum~\cite{Gasperini:1993yf}. 
In this framework, the expectation value of the number density operator is 
determined by the Bose-Einstein distribution. {}For a gauge field in a thermal 
bath at physical temperature $\mathcal{T}$, the expectation value for the
number density is given by
\begin{equation}\label{Thermal_PS}
    \langle \Omega | a^\dagger_{\mathbf{k},\lambda} a_{\mathbf{k}',\lambda'} 
|\Omega\rangle = (2\pi)^3 \delta^3(\mathbf{k} - \mathbf{k}') 
\delta_{\lambda\lambda'} \left( \frac{1}{e^{k/(a\mathcal{T})} - 1} \right)\,,
\end{equation}
where $k/a$ represents the physical momentum. Using this occupation number, 
we compute the thermally-modified spectral magnetic energy density. By 
substituting \eqref{Thermal_PS} into the standard expression for the magnetic 
power spectrum \eqref{magnetic_power_spectrum}, we obtain that
\begin{equation}
    \frac{d \rho_B}{d \ln k} = \frac{k^4}{4\pi^2 a^4} 
\coth\left(\frac{k}{2 a \mathcal{T}}\right)\,.
\end{equation}

\begin{figure}[!htb]
    \centering
    \includegraphics[width=11cm]{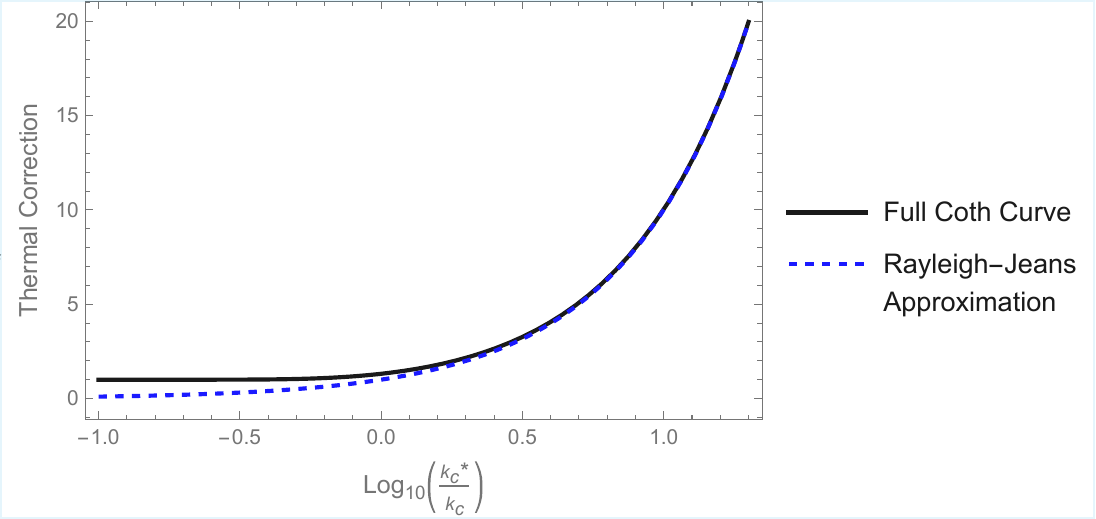}
    \caption{Thermal correction factor $\coth\!\left(k/2a\mathcal{T}\right)$ compared with its Rayleigh-Jeans approximation,  plotted as a function of the dimensionless ratio $\log_{10}(k_c^* / k_c)$. $\lambda^*_c=2\pi / k_c^*$ is the comoving pivot scale introduced satisfying $k_c^*/a\simeq 2\mathcal{T}$. The temperature here is chosen to be $\mathcal{T}=10^{10}\rm GeV$ for a specific illustration.}
    \label{fig:coth_plot}
\end{figure}

The term $\coth(k/2 a \mathcal{T}) \equiv 1+ 2 n_k$ encapsulates the departure from 
the vacuum state due to thermal fluctuations. In the limit $\mathcal{T} \to 0$, 
this factor approaches unity, effectively recovering the standard Bunch-Davies 
vacuum result. Conversely, the thermal enhancement is maximized in the infrared (IR) regime, where the physical momentum is much smaller than the temperature, 
$(k/a) \ll \mathcal{T}$. In this Rayleigh-Jeans limit, the modification factor approximates to
\begin{equation}
    \coth \left(\frac{k}{2 a \mathcal{T}}\right) \approx \frac{2 a \mathcal{T}}{k}\,,
\end{equation}
confirming that the amplification is most significant for long-wavelength modes and 
high physical temperatures. See Fig.\ref{fig:coth_plot} for a demonstration of the validity of this approximation in this limit.

To relate these findings to observable magnetic fields, we evaluate the power spectrum in the super-horizon limit ($k \ll aH$):
\begin{equation}\label{Thermal_PS_full}
    \frac{d \rho_B}{d \ln k} \approx \frac{k^3 \mathcal{T}}{2\pi^2 a^3}\,,
\end{equation}
where we maintain the physical temperature $\mathcal{T}$ as a constant during 
the inflationary phase. A crucial observation here is the scaling behaviour: while 
the vacuum magnetic energy density dilutes as $a^{-4}$ (standard conformal radiation scaling), in the presence of a thermalized heat bath of photons there is a {\it dissipative boost} which leads to the spectrum decaying only as $a^{-3}$. This slower dilution results in a cumulative enhancement factor relative to the vacuum case, which by the present epoch $\tau_0$ is given by
\begin{equation}
    \chi (\tau_{\rm end}) = \frac{a(\tau_0)}{a(\tau_{\rm end})}\,,
\end{equation}
where $a(\tau_{\rm end})$ is the scale factor at the end of inflation.

It is important to emphasize once again that conformal invariance remains intact at the level of the action. The conformal symmetry breaking required for magnetogenesis is achieved here solely through the choice of the initial state, characterized by the preferred physical scale $\mathcal{T}$. This state-dependent conformal symmetry breaking implies that while sub-horizon
modes with $k/a \gg \mathcal{T}$ remain largely unaffected and behave like
standard vacuum fluctuations, modes satisfying $k/a \lesssim \mathcal{T}$ can
receive an enhanced occupation relative to the Bunch-Davies case. Under the
maintained-thermal-state prescription discussed below, this enhancement can be
large enough for the resulting magnetic fields to potentially serve as
primordial seeds.

A qualification is nevertheless important for the long-wavelength modes. What we assume is the existence of a thermal state for all modes without going into a discussion of how they can possibly become thermal. For modes well inside the Hubble radius, the use of a thermal occupation number is physically well motivated, since local interactions with the bath can maintain approximate equilibrium. For modes with $k/a\ll H$, however, this
interpretation is less direct: a Fourier mode is a nonlocal object and cannot, in general, be assumed to remain in causal thermal contact with the bath\footnote{In some models of preheating and reheating for example, non-causal amplification for superhorizon curvature modes exists as a result of parametric resonance \cite{Liddle:1999hq,Bassett:2005xm}. This is different from our mechanism during the pseudo de Sitter phase, but illustrates
a physical situation where the amplification of superhorizon modes is nontrivial in principle.}. In the
absence of an explicit microscopic calculation showing that the relevant dissipation and noise kernels continue to act efficiently on such modes, a conservative prescription is to evaluate the thermal occupation at horizon crossing and then keep it fixed. In that case, after horizon exit the magnetic energy density redshifts as $a^{-4}$, as in the Bunch-Davies case, but with a finite thermal enhancement,
\begin{equation}
\label{subhorizon}
\rho_B \simeq \rho_{B,{\rm vac}}\left[1+2n_k(t_k)\right]
       \simeq \rho_{B,{\rm vac}}\,\frac{2{\cal T}_k}{H_k},    
\end{equation}
where $t_k$ denotes the time of horizon crossing and we have assumed
${\cal T}_k\gg H_k$. The alternative scaling $\rho_B\propto a^{-3}$ corresponds to
the stronger phenomenological assumption that the gauge sector remains an open
subsystem in an approximately thermal state even for the long-wavelength modes,
so that the occupation number continues to track the instantaneous thermal
factor $n_k(t)\simeq a{\cal T}/k$. The results below should be understood in this
second, maintained-thermal-state sense.

An immediate consequence is that our setup, at least in principle, allows us to evade the Green-Kobayashi no-go result which assumes quantum fluctuations around the Bunch-Davies vacuum\footnote{Even the case of classical growth discussed in \cite{2016JCAP...03..010G} assumes a stochastic average of the form $\langle a^\dagger_{{\bf k},\lambda} a_{{\bf k'},\lambda'} \rangle = \left(2\pi\right)^3 \delta^3({\bf k} - {\bf k'}) \ \delta_{\lambda\lambda'}$ which specifically misses the thermal distribution function in \eqref{Thermal_PS}.}. Here we replace vacuum initial conditions with thermal
ones, which alters the spectrum of the gauge field, Eq.~\eqref{Thermal_PS_full}.
In the maintained-thermal-state prescription adopted below, the physical
temperature of the bath remains approximately constant during inflation due to
dissipative effects, and the thermal occupation number continues to scale as
$n_k\simeq a{\cal T}/k$. This is what makes the red-shifting of the final result
weaker by one power of the scale factor, namely $\rho_B\propto a^{-3}$ rather
than $a^{-4}$.

\subsection{Evolution and Present-Day Field Amplitudes}

Inflation is followed by a reheating phase, the canonical picture for which is that the inflaton field oscillates around some minima to decay into the Standard Model particles populating the universe. This creates a high-conductivity plasma during which phase, the ideal magnetohydrodynamic (MHD) condition applies, \textit{i.e.,} $\mathbf{E} = -\mathbf{v} \times \mathbf{B}$ to leading order. Since the magnetic flux is conserved during this phase,  $\Phi_B = {\rm const.}$, the magnetic field evolves as $B\propto1/a^2$ (in physical coordinates) irrespective of the initial conditions. We shall not modify this condition in this work since we had only assumed that the physical temperature remains constant during inflation, and then decays as usual during reheating. However, it is worth pointing out that this conclusion will change drastically if we were to assume a full warm inflationary scenario instead of only assuming a transiently thermal state. In other words, our thermal modification is only limited to the inflationary phase and further corrections would come from warm inflation where a continuous feeding of the radiation bath replaces standard reheating. This will be pursued in future work. 

For the given model, to make connections with observations, we need to define the spectral magnetic field amplitude at comoving wavelength $\lambda = 2\pi/k$ as
\begin{eqnarray}\label{sqrt}
	B_\lambda = \sqrt{2 \frac{{\rm d} \rho_B}{{\rm d} \ln k}}\,,
\end{eqnarray}
which gives an order of magnitude (the RMS value) of the magnetic field smoothed over the length scale $\lambda$. Hence, the characteristic coherence length of this mode is given by the inverse of the wavenumber, $\lambda_c(\tau) \sim 1/(a H_{\rm inf})$. This means that modes which exit the horizon during inflation at Hubble scale $H_{\text{inf}}$ maintain coherence scales of order megaparsecs or larger by the present epoch, providing a natural mechanism for large-scale magnetic field correlations observed at late-times. Similarly, the spectral index on super-horizon scales, for the thermal state, can be computed from \eqref{Thermal_PS_full} as $n^B_{\rm thermal} = 3$, as opposed to $n^B_{\rm vac} =4$ in the vacuum case. After inflation, MHD turbulence and viscous damping can steepen the spectrum on small scales while preserving the large-scale structure fixed by the thermal spectrum \cite{Subramanian:2015lua, Durrer:2013pga, Kahniashvili:2010gp}.

Considering inflation is followed by reheating, one can introduce the following parameterization for the number of $e$-foldings a Fourier mode of comoving wavelength $\lambda$ spent outside the horizon before the end of inflation (where we stick to the notation of \cite{Turner:1987bw}):
\begin{eqnarray}\label{Turner}
	N = 45 + \ln\left(\frac{\lambda}{{\rm Mpc}}\right) + \frac{2}{3} \ln\left(\frac{M_{\rm inf}}{10^{14} {\rm GeV}}\right) + \frac{1}{3}\ln\left(\frac{\mathcal{T}_{\rm reh}}{10^{10} {\rm GeV}}\right)\,,
\end{eqnarray}
when we set the scale factor today as $a_0=1$. Here, $M_{\rm inf}$ is the inflation energy scale given by $M_{\rm inf}^2 = \sqrt{3}H_{\rm inf}M_{\rm Pl}$.

The dimensionless magnetic field energy ratio at the end of inflation is given by
\begin{eqnarray}
	r^{\lambda} = \frac{\rho_\lambda^{(B)} ({\rm end})}{\rho_{\rm total} ({\rm end})}\,.
\end{eqnarray}
Using the condition for horizon crossing $k=aH$, one can find the value of this energy ratio for the standard vacuum case as 
\begin{eqnarray}\label{r_vac}
	r_{\rm vac} \simeq\left(\frac{H_{\rm inf}}{M_{\rm Pl}}\right)^2e^{-4N}\,,
\end{eqnarray}
assuming that superhorizon modes can be treated classically. It is straightforward to see from \eqref{Turner} and \eqref{r_vac} that the field-amplitude scales as $B_{\rm vac}\propto\left(\frac{M_{\rm inf}}{\mathcal{T}_{\rm reh}}\right)^{2/3}$. Conversely, for the thermal state, one finds
\begin{eqnarray}\label{therm}
	r_{\rm therm} \simeq \left(\frac{H_{\rm inf} \mathcal{T}}{M^2_{\rm Pl}}\right) \, e^{-3N}\,.
\end{eqnarray}
Keeping a warm-inflationary scenario in mind, it is imperative to set the temperature of the bath to be equal to that of reheating at the end of inflation. In other words, once inflation ends, the energy density of the photon bath dominates the energy budget of the whole cosmos. Under this warm inflation constraint $\mathcal{T}\sim \mathcal{T}_{\rm reh}$, one easily deduces from \eqref{Turner} and \eqref{therm} that both the energy density ratio, and hence the magnetic field, are now independent of parameter choices since $B_{\rm therm}, r_{\rm therm} \propto \left(\frac{M_{\rm inf}}{\mathcal{T}_{\rm reh}}\right)^{0}$. 
Hence, the thermodynamic boost achieved in our mechanism relative to the vacuum result is
\begin{equation}\label{boost}
    \frac{r_{\rm therm}}{r_{\rm vac}} = \frac{{\cal T}}{H_{\rm inf}} e^N.
\end{equation}
In arriving at the above estimates, we have used an instantaneous reheating approximation. The magnetic energy ratio is shown to receive a potential boost given the strong thermal assumption, \textit{i.e.} $\frac{\mathcal{T}}{H_{\rm inf}}\gtrsim 1$. In particular, we note that in most well-motivated warm inflation models (see, e.g.,
refs.~\cite{Bastero-Gil:2019gao,ORamos:2025uqs}), it is quite easy to
reach values of $\frac{\mathcal{T}}{H_{\rm inf}}\gtrsim 10^3$, hence
representing quite a significant enhancement. The other crucial thing to note is the reduced decay rate of the superhorizon modes reflected by the dependence $e^{-3N}$ rather than $e^{-4N}$ for the vacuum case, which is independent of the value of $\mathcal{T}$.

 To have an estimate of the effects of the \textit{dissipative boost}, let us set, for example, $M_{\rm inf}=\mathcal{T}_{\rm reh}=10^{10}{\rm GeV}$. In tables~\ref{table1} and \ref{table2} we present the results varying each of $M_{\rm inf}$ and $\mathcal{T}_{\rm reh}$ choosing some representative values. Following the conventional scaling, the strength of a primordial magnetic field at the end of inflation is estimated to be $10^{-50} \rm G$ in the standard vacuum case. We then deduce that the magnetic field at a coherence scale of $\sim 10{\rm kpc}$ relevant for galactic dynamos \cite{Widrow:2002ud}, has a present day strength of $10^{-53}\text{G}$, assuming $r_{\rm vac}$ stays constant after inflation. A handy relation is the radiation energy density today ($t=t_0$) given in Gauss units \cite{Durrer:2013pga}:
\begin{equation}
	\rho_{\rm rad}(t_0)\approx\frac{(3\times10^{-6}G)^2}{8\pi}\,.
\end{equation}
However, to be astrophysically significant on this scale, a seed field has to be $\sim 3 \times 10^{-19}{\rm G}$ to be meaningfully amplified by galactic dynamos \cite{Turner:1987bw}, implying an unfilled gap of $10^{34}$ between what observations demand and standard vacuum predictions.

Keeping the parameter choices as before, and now considering a thermal bath with temperature around $\mathcal{T}=10^{10} {\rm GeV}$, we find
\begin{equation}
    r^{\lambda=10 {\rm kpc}}_{\rm therm} = 10^{-72}=10^{24} \times r^{\lambda=10 {\rm kpc}}_{\rm vac}\,.
\end{equation}
This corresponds to a cosmological magnetic field, that is coherent across lengths $\lambda=10 {\rm kpc}$, with a present-day strength of $B_0=10^{-41}G$. A typical temperature range for warm inflation is given by $10^6 - 10^{12}{\rm GeV}$. Recall that the amplification is directly proportional to the thermal bath temperature given \eqref{therm}.

We can repeat the same analysis for comoving wavelength $\lambda=1{\rm Mpc}$ as follows:
\begin{equation}
    r^{\lambda = 1{\rm Mpc}}_{\rm therm}= 10^{-78}=10^{26} \times r^{\lambda = 1{\rm Mpc}}_{\rm vac}\,.
\end{equation}
Despite leaving a remaining gap of $10^{28}$ between the thermally amplified prediction $B=10^{-44}{\rm G}$ and the lower bound for cosmic magnetic fields $B\gtrsim 10^{-16}{\rm G}$ at $\lambda=1{\rm Mpc}$ \cite{Neronov:2010gir}, it is evident that the thermal correction significantly brings down the gap. The upper bound for the present-day strength of primordial magnetic fields is around the order of nano Gauss $(10^{-9}\text{G})$ at the scale of 1 Mpc in \cite{Planck:2015zrl}. For all reasonably large astrophysical and cosmological scales, such amplifications are expected to be present. As shown above, thermal occupation of lower-$k$ modes naturally shifts power towards larger comoving scales.

\begin{table}[!htpb]
      \caption{Comparison of present-day magnetic field strengths ($B_0$) at a comoving scale of $\lambda = 1$ Mpc for
  various reheating temperature choices, keeping the inflation scale fixed at $M_{\rm inf}=10^{10}$
    GeV. We set the thermal bath temperature to $\mathcal{T}\sim \mathcal{T}_{\rm reh}$ here for this table, restricting
    to benchmark values safely below the inflation scale. Note that the dissipative boost  $r_{\rm therm}/r_{\rm vac}$
  is with respect to the magnetic field energy $\propto B_0^2$.}\medskip
          \label{table1}
          \centering
      \small
      \begin{tabular}{@{}lcccc@{}}
      \toprule
      \textbf{Reheating Temp.} ($\mathcal{T}_{\rm reh}$) &
       $B_0^\textbf{vac}$ &
       $B_0^\textbf{therm}$ &
     $r_{\rm therm}/r_{\rm vac}$ \\
      \midrule
      High ($10^{9}$ GeV)         & $\sim 10^{-56}$ G & $\sim 10^{-44}$ G & $\sim 10^{24}$ \\
      Intermediate ($10^{6}$ GeV) & $\sim 10^{-54}$ G & $\sim 10^{-44}$ G & $\sim 10^{20}$ \\
      Low ($10^{2}$ GeV)          & $\sim 10^{-51}$ G & $\sim 10^{-44}$ G & $\sim 10^{15}$ \\
      \midrule
      \end{tabular}
    \end{table}

    \begin{table}[!htpb]
      \caption{Similar to table~\ref{table1}, but now keeping fixed the reheating temperature at $\mathcal{T}_{\rm reh}
    =10^{9}$ GeV and varying the inflation scale $M_{\rm inf}$. We again set the thermal bath temperature to $
  \mathcal{T}\sim \mathcal{T}_{\rm reh}$ and only display benchmark points safely above the reheating
  temperature.}\medskip
          \label{table2}
          \centering
      \small
      \begin{tabular}{@{}lcccc@{}}
      \toprule
      \textbf{Inflation Scale} ($M_{\rm inf}$) &
     $B_0^\textbf{vac}$ &
       $B_0^\textbf{therm}$ &
      $r_{\rm therm}/r_{\rm vac}$ \\
      \midrule
      High ($10^{16}$ GeV)         &  $\sim 10^{-52}$ G & $\sim 10^{-44}$ G & $\sim 10^{16}$ \\
      Intermediate ($10^{13}$ GeV) &  $\sim 10^{-54}$ G & $\sim 10^{-44}$ G & $\sim 10^{20}$ \\
        Low ($10^{11}$ GeV)          &  $\sim 10^{-55}$ G & $\sim 10^{-44}$ G & $\sim 10^{23}$ \\
      \midrule
      \end{tabular}
    \end{table}

The quantitative impact of the modified redshift scaling is summarized in Table~\ref{table1} for modes at $\lambda=1\rm Mpc$, where we compare the present-day magnetic field strength $B_0^{\rm vac}$ for some representative reheating temperature scales while setting the inflation scale to be $10^{10}$ GeV. The vacuum predictions scale as $B_0\propto\left(M_{\rm inf}/\mathcal{T}_{\rm reh}\right)^{2/3}$, whereas the thermally corrected field amplitude is effectively independent of both parameters in the present setup, subject to the warm inflation condition $\mathcal{T}\sim \mathcal{T}_{\rm reh}$. The dissipative boost shown in the table corresponds to the ratio of
  magnetic energy densities, and therefore scales as
  \begin{eqnarray}
  \frac{r_{\rm therm}}{r_{\rm vac}}\propto \left(\frac{\mathcal{T}_{\rm reh}}{M_{\rm inf}}\right)^{4/3}\,,    
  \end{eqnarray}
while the enhancement in the field amplitude $B$, eq.~\eqref{sqrt}, is its square root. Table \ref{table2} shows the corresponding results for different $M_{\rm inf}$ values while holding the reheating temperature fixed. In particular, a standard GUT-scale inflation ($M_{\rm inf} \approx 10^{16}$~GeV), the vacuum prediction for the seed field is approximately $10^{-52}$~G, a value far too suppressed to match observations. In contrast, the thermal state -- by virtue of the $a^{-3}$ energy density scaling -- provides a robust boost of sixteen orders of magnitude, reaching $\sim 10^{-44}$~G. While this amplified field remain below the $B_0 \gtrsim 10^{-16}$~G lower bound inferred from the non-observation of GeV gamma-ray cascades from TeV blazars \cite{2010Sci...328...73N, Tavecchio:2010mk, Neronov:2010gir}, the trend across different energy scales is instructive. Even when the inflation scale and reheating temperature gets varied, the dissipative boost, and the resulting absolute field strength, remains consistently significant. Similarly, at the galactic coherence scale $\lambda=10~{\rm kpc}$, the corresponding thermal enhancement is from $10^{13}$ to $10^{22}$, or equivalently $B_0^{\rm therm}/B_0^{\rm vac}\sim 10^7$--$10^{11}$ in field amplitude. These results demonstrate that while thermal conditions substantially alleviate the conformal obstruction, the isolated assumption of a fixed temperature in a cold background is insufficient. This strongly motivates a transition to a fully dynamical warm inflation scenario, where continuous energy dissipation from the 
inflaton into the radiation bath may provide the additional amplification 
required to satisfy these stringent observational constraints.

Another point to note here is that the electric field will also get a similar $\coth{(k/a\mathcal{T})}$ factor due to the thermal state. This can lead to an electric field that is stronger than the magnetic field. If reheating is not instant and there is a low-conductivity phase following inflation, this will also induce a magnetic field  according to the relation $|B|^2 \sim 1/\left(a^6H^2\right)$  \cite{Kobayashi:2019uqs}. For matter-dominated expansion, this leads to a further enhancement in the magnetic field spectrum as it decays as $|B|^2\sim 1/a^3$, instead of going as $1/a^4$,  \textit{post inflation} as well. This will also be explored in future work when considering a smooth transition from warm inflation to a radiation-dominated era without the need for reheating.

\section{Conclusions}
\label{sec4}

The origin of large-scale primordial magnetic fields  remains an important
unresolved question in modern cosmology. While inflation provides a natural 
framework for generating super-horizon correlations, producing phenomenologically viable seed fields without violating basic consistency requirements -- such as back-reaction limits or strong coupling -- remains a significant challenge. The primary obstruction is the conformal invariance of standard Maxwell electrodynamics in a flat FLRW background, which causes the magnetic energy density to decay as $1/a^4$, a rate too rapid to seed the fields observed in galaxies and the intergalactic medium today. Consequently, most models rely on explicit breaking of the conformal symmetry through non-minimal couplings of the gauge field to the inflaton or dark matter \cite{Kamali:2026tgq}, or by significantly altering the 
background dynamics \cite{Atkins:2025pvg}. Notable exceptions, such as the quantum conformal anomaly~\cite{Campanelli:2013mea} or spatially curved backgrounds~\cite{Barrow:2012ax}, come with their own potential theoretical 
pathologies~\cite{Durrer:2013xla,Shtanov:2012pp}.

In this work, we have explored a different trajectory: maintaining the minimal action of standard electromagnetism, while modifying the initial state of the gauge field. By assuming the gauge field is prepared in a thermal state---a condition naturally motivated by the environment of warm inflation---we demonstrate a substantial enhancement in the primordial magnetic spectrum. The fundamental physical mechanism driving this \textit{dissipative boost} is the physical 
temperature $\mathcal{T}$, which does not redshift in the standard $1/a$ fashion, due to a sustained thermal bath. This effectively leads to a decay of the magnetic energy density as $1/a^3$ instead of $1/a^4$. Hence, having a thermal bath of gauge particles with a constant physical temperature is central to our mechanism. Whether this is justifiable for an inflationary background will be examined in the future within a robust warm-inflation scenario. 

Even if conditions for superhorizon thermalization to happen depend on the details of the mode dynamics treated as an open quantum subsystem, a thermal enhancement of the magnetic fields within the horizon during warm inflation in \eqref{subhorizon} has already the potential to leave phenomenological signatures as a consequence of the modified initial condition for later plasma evolution through, for example, anisotropic stress and Lorentz force. These thermally produced magnetics fields can also leave well-known signatures on CMB temperature and polarization anisotropies through scalar, vector and tensor perturbations~\cite{Planck:2015zrl}. For a stochastic source of primordial magnetic field in particular, such anisotropy contribution will be non-Gaussian, and thus, has the potential of affecting the CMB bispectrum~\cite{Finelli:2018uwp}.  On sufficiently small scales, magnetic field induced Lorentz forces can also generate extra baryon perturbations, with possible consequences for early structure formation~\cite{Ralegankar:2023vyn}, reionization and 21-cm observables~\cite{Tashiro:2006uv}. The quantitative impact depends on the properties of the magnetic fields such as 
helicity~\cite{Long:2013tha} and post-inflationary plasma history and is therefore left for future work.

This thermal occupation of long-wavelength modes slows the effective decay of super-horizon magnetic energy, yielding present-day magnetic field amplitudes $B_0$, that can be potentially about $10^{8}$ to $10^{12}$ times larger than those obtained from the standard Bunch-Davies vacuum on cosmological scales $\sim 1\rm Mpc$ and correspondingly $10^7$ to $10^{11}$ times larger on galactic scales $\sim 10 \rm kpc$. The overall value is only very weakly dependent on the ratio between the scale of inflation and the reheating temperature at the start of the radiation dominated regime after inflation (see Table~\ref{table1} and Table~\ref{table2} for representative examples).  Critically, this amplification is achieved without invoking non-minimal couplings in the electromagnetic sector. While our simplified model assumes a cold inflationary background followed by instantaneous reheating to establish predictions, we emphasize that this serves as an illustrative toy model for a broader out-of-equilibrium quantum field theory approach. In a more general setting, the interaction of gauge photons with environmental fields (such as fermions) would lead to a mixed state for the electromagnetic field, providing a more robust physical foundation than a simple thermal assumption while leading to a more intricate relation for the dissipative boost.

Despite the significant enhancement reported here, the predicted magnetic field amplitudes still fall below the strengths inferred from observations of galaxies, clusters, and the intergalactic medium. Our results suggest that thermal initial conditions, while highly beneficial, may not be sufficient on their own to resolve the magnetogenesis problem within a purely cold-inflationary framework. Furthermore, maintaining a fixed physical temperature during quasi-de Sitter expansion is not a `natural' state in isolation, as it requires the continuous production of radiation. This is precisely where the paradigm of warm inflation becomes essential. A natural next step is to embed this mechanism into a fully dynamical warm inflation setup, where the inflaton, radiation bath, and electromagnetic field interact consistently throughout the inflationary expansion. Such a framework would eliminate the need for a discrete reheating phase and could potentially lead to even stronger magnetic fields by accounting for continuous energy 
dissipation into the gauge sector. {}Furthermore, this would alleviate the need to have a high conductivity phase right after inflation, by eliminating instant reheating, thereby opening the possibility of inducing further magnetic fields due to strong electric fields. We leave the detailed analysis of this complete dynamical scenario for future work.


\section*{Acknowledgments}
S.B. thanks Arko Bhaumik for useful comments on an earlier version of this draft. 
A.B. is partially supported by STFC. S.B. is supported in part by a start-up grant from the Indian Statistical Institute. 
R.O.R. acknowledges financial support by research grants from Conselho
Nacional de Desenvolvimento Cient\'{\i}fico e Tecnol\'ogico (CNPq),
Grant No. 307286/2021-5, and from Funda\c{c}\~ao Carlos Chagas Filho
de Amparo \`a Pesquisa do Estado do Rio de Janeiro (FAPERJ), Grant
No. E-26/200.415/2026.



\end{document}